\documentclass[aps,floats,twocolumn,prd,nofootinbib,superscriptaddress]{revtex4-2}

\usepackage{fnpct} 
\usepackage{comment}
\usepackage{ifpdf}
\usepackage{slashed}
\usepackage{color}
\usepackage[mathscr]{eucal}
\usepackage[utf8]{inputenc}
\usepackage{cancel}
\usepackage{bm}

\usepackage{braket}

\usepackage[most]{tcolorbox}

\tcbset{colback=white, colframe=black, 
        highlight math style= {enhanced, %<-- needed for the ’remember’ options
            colframe=red,colback=red!10!white,boxsep=0pt}
        }

\ifpdf
  \usepackage{graphicx}     %   usepackage without driver option
  \usepackage[bookmarksopen,colorlinks=true,linkcolor=bblue,citecolor=bblue,urlcolor=ppink]{hyperref}
\else     % For (p)LaTeX + dvipdfmx
\fi

\definecolor{red}{rgb}{1,0,0}
\definecolor{darkred}{rgb}{0.6,0,0}
\definecolor{darkgreen}{rgb}{0.992447,0.623778,0.034597}
\definecolor{ppink}{rgb}{1,0.4,0.4}
\definecolor{bblue}{rgb}{0.284602,0.317763,0.963947}
\definecolor{purple}{rgb}{0.5 ,0, 0.7}

\definecolor{dgreen}{rgb}{0 ,0.5, 0.5}

\newcommand{\vev}[1]{ \left< {#1} \right> }
\newcommand{\dd}{\mathrm{d}}

\newcommand{\ee}{\text{e}}

\makeatletter
\newcommand\footnoteref[1]{\protected@xdef\@thefnmark{\ref{#1}}\@footnotemark}
\makeatother

\allowdisplaybreaks[1]

\begin{document}

\title{
Adding electromagnetic birefringence to pulsar timing and astrometry to detect gravitational waves}

\author{Keisuke Inomata}
\email{inomata@jhu.edu}
\affiliation{
 William H.\ Miller III Department of Physics \& Astronomy, Johns Hopkins University, 3400 N.\ Charles St., Baltimore, MD 21218, USA
}

\author{Marc Kamionkowski}
\email{kamion@jhu.edu}
\affiliation{
 William H.\ Miller III Department of Physics \& Astronomy, Johns Hopkins University, 3400 N.\ Charles St., Baltimore, MD 21218, USA
}

\begin{abstract}
It was recently shown that the time variation of the polarization of electromagnetic waves from pulsars can be used, in cross-correlation with pulsar timing, to probe the chirality of an isotropic gravitational wave background.  Here, we show that the expression for the cross-correlation is derived efficiently with the total-angular-momentum formalism and use this framework to extend the formulation to cross-correlation with astrometry.  We do so for spin-1 gravitational waves (that may arise in alternative-gravity theories) as well as the general-relativistic spin-2 gravitational waves.
\end{abstract}

\maketitle

\section{Introduction}
The possible origins of a nano-Hertz stochastic gravitational-wave background (SGWB) have been actively investigated since evidence from pulsar-timing arrays (PTAs) \cite{NANOGrav:2023gor,EPTA:2023fyk,Reardon:2023gzh,Xu:2023wog} for its detection was recently presented.  While the SGWB is usually presumed to have no handedness (i.e., equal amplitudes of right- and left-handed modes), there may be reasons from gravitational parity violation---e.g., from Chern--Simons \cite{Lue:1998mq, Alexander:2004us, Satoh:2007gn, Alexander:2009tp, Kanno:2023kdi} or Ho\v{r}ava--Lifshitz gravity~\cite{Takahashi:2009wc,Wang:2012fi,Zhu:2013fja}---or from parity violation in the matter sector~\cite{Sorbo:2011rz,Cook:2011hg,Dimastrogiovanni:2012ew,Adshead:2013nka, Dimastrogiovanni:2016fuu,Obata:2016tmo,Adshead:2018doq,Bastero-Gil:2022fme,Aoki:2025uwz} for the SGWB to be chiral.  However, a PTA is insensitive to the GW chirality \cite{Kato:2015bye,Belgacem:2020nda} (although circular-polarization {\it anisotropies} are detectable \cite{Kato:2015bye,Belgacem:2020nda,Sato-Polito:2021efu}), while it can be detected through cross-correlation with an astrometry survey~\cite{Qin:2018yhy,Liang:2023pbj,Caliskan:2023cqm,Inomata:2024kzr}.

An interesting recent paper \cite{Liang:2025vji}  has proposed another possibility: cross-correlation of pulsar timing residuals with the rotation of the linear polarization of the electromagnetic waves from the pulsar.  This rotation is a pseudoscalar, and so its cross-correlation (``pulsar polarimetry'') with the pulse arrival time (a scalar on the celestial sphere) probes chirality.

Here we show that the rotation-timing cross-correlation derived in Ref.~\cite{Liang:2025vji} follows very naturally from the total-angular-momentum (TAM) formalism \cite{Dai:2012bc}, where GW plane waves (eigenstates of linear momentum) are replaced by TAM waves (eigenstates of angular momentum). The TAM formalism also allows us to quickly obtain the cross-correlation with astrometry observables and to generalize the results to spin-1 GWs.

\section{Analysis}

\subsection{TAM waves}

Consider the contribution,
\begin{equation}
     h_{ab}(\boldsymbol{x},t) = 4\pi i^\ell h^{k,X}_{(\ell m)}
     \Psi^{k,X}_{(\ell m)ab}(\boldsymbol{x}) \ee^{-ikt},
     \label{eq:tam_mode}
\end{equation}
of a single TAM wave $\Psi^{k,X}_{ab}$, with $X \in \{TE,TB,VE,VB\}$.\footnote{We do not consider the scalar-longitudinal/transverse TAM mode ($SL,ST$) because they do not break parity.}  Here, TE and TB are respectively parity even and odd TAM polarizations for spin-2 GWs, and VE and VB are the analogs for spin-1 GWs.  
We introduce radial function $R^{p,X}_\ell$ \cite{Qin:2018yhy} through
\begin{align}
    \hat n^a \Psi^{X}_{(\ell m)ab} &= R^{L,X}_\ell(kr) Y^L_{(\ell m) b}(\hat{\bm n}) + R^{E,X}_\ell(kr) Y^E_{(\ell m) b}(\hat{\bm n}) \nonumber \\
    &\quad + R^{B,X}_\ell(kr) Y^B_{(\ell m) b}(\hat{\bm n}),
\end{align}
where $\bm{\hat n}$ is the line-of-sight unit vector ($\bm x = r \bm{\hat n}$), and $Y^{p}_{(\ell m)a}$ are vector spherical harmonics in the longitudinal/transverse basis~\cite{Dai:2012bc}.  The radial eigenfunctions are \cite{Qin:2018yhy},
\begin{align}
 &R^{L,TE} = -N_\ell \frac{j_\ell(kr)}{(kr)^2}, \qquad R^{L,TB} = R^{L,VB} = 0,
 \\
 &R^{L,VE} = - \frac{\sqrt{2\ell(\ell+1)}}{kr} \left( j'_\ell(kr) - \frac{j_\ell(kr)}{kr} \right),
\end{align}
where $N_{\ell} = \sqrt{(\ell+2)!/[2(\ell-2)!]}$ and $j_\ell(x)$ is the spherical Bessel function of the first kind.

\subsection{Timing residuals}

The redshift induced by the metric perturbation is 
(e.g., Eq.~(23.10) in Ref.~\cite{Maggiore:2018sht}),
\begin{equation}
     z(\boldsymbol{\hat{n}},t) = \frac12 
     \int_{t-r_s}^t \, \dd t'\,
     \frac{\partial}{\partial t'} \hat n^a \hat n^b h_{ab}\left(\boldsymbol{x}(t'),t'
     \right),
     \label{eq:z_1}
\end{equation}
where $r_s$ is the distance to the pulsar (or the star for astrometry) and $\partial/\partial t'$ acts only on the first argument ($t'$), not the second ($\boldsymbol{x}(t')$).  
Note that we have already neglected the pulsar terms that do not contribute to the two-point correlation.
From Eq.~(\ref{eq:z_1}), we obtain,
\begin{equation}
     z(\boldsymbol{\hat{n}},t) = -\frac{i}{2}
     4\pi  i^\ell h^{k,X}_{(\ell m)}
     Y_{(\ell m)}(\boldsymbol{\hat{n}}) \ee^{-ikt}
     \int_0^{k r_s}\, \dd x\, R_\ell^{L,X}(x) \ee^{ix},
\label{eq:z_2}
\end{equation}
which in the distant-source limit $kr_s \to \infty$ becomes
\begin{equation}
    \label{eq:z_final}
     z(\boldsymbol{\hat{n}},t) = 4\pi i^\ell F_\ell^{z,X} h^{k,X}_{(\ell m)} Y_{(\ell m)}(\boldsymbol{\hat{n}}) \ee^{-ikt},
\end{equation}
where $F^{z,X}_\ell$ is given in Table~\ref{tab:F}.  Note that only the TE/VE modes generate a timing residual, not the TB/VB modes.

\subsection{Electromagnetic Birefringence}

Eq.~(66) in Ref.~\cite{Liang:2025vji} derives the angle $\psi$ by which the linear polarization of an electromagnetic wave is rotated, while it was also derived earlier in  Ref.~\cite{Dai:2013nda}.  It is
\begin{equation}
\label{eq:psi}
\psi({\boldsymbol{\hat n}},t) = -\frac{1}{2} \varepsilon_{abc} \hat n^c \int^{t}_{t-r_s} \dd t'\, \hat n^d \partial^a h^{bd}(\bm x(t'),t'),
\end{equation}
where $\varepsilon_{abc}$ is the anti-symmetric Levi-Civita tensor.
For a single TAM wave, this becomes
\begin{align}
    \label{eq:psi_tam}
    \psi(\boldsymbol{\hat{n}},t) &= -\frac{i}{2}
     4\pi  i^\ell h^{k,X}_{(\ell m)}
     Y_{(\ell m)}(\boldsymbol{\hat{n}}) \ee^{-ikt} \nonumber \\ 
     & \quad \times 
     \int_0^{k r_s}\, \dd x\, \frac{\sqrt{\ell(\ell + 1)}}{i x} R_\ell^{B,X}(x) \ee^{ix},
\end{align}
where\footnote{The Appendix derives a useful relation between the derivative of $TB/VB$ TAM modes and the spherical harmonics.}
\begin{align}
 & \frac{\sqrt{\ell(\ell + 1)}}{i x} R_\ell^{B,TB}(x) = R^{L,TE}(x), \nonumber \\
 & \frac{\sqrt{\ell(\ell + 1)}}{i x} R_\ell^{B,VB}(x) = \frac{1}{2}R^{L,VE}(x),
\end{align}
and $R^{B,TE} = R^{B,VE} = 0$.
We thus find
\begin{equation}
    \psi(\boldsymbol{\hat{n}},t) = 4\pi i^\ell F_\ell^{\psi,X} h^{k,X}_{(\ell m)} Y_{(\ell m)}(\boldsymbol{\hat{n}}) \ee^{-ikt},
     \label{eq:psi_final}
\end{equation}
where $F^{\psi,X}_\ell$ is given in Table~\ref{tab:F}.  Note that this vanishes for the TE/VE mode but not for the TB/VB.

\begin{table*}
\centering
    \everymath{\displaystyle}
    \begin{tabular}{|c|c|c|c|c|}
    \hline
  $X$ & $F^{z,X}_\ell$ & $F^{\psi,X}_\ell$ & $F^{{E},X}_\ell$ & $F^{{B},X}_\ell$  \\
  \hline \hline
  $VE$ & $ -\frac{i}{3} \delta_{\ell1} + \frac{i^\ell}{\sqrt{2\ell (\ell+1)}}$ & 0 & 
  $\frac{2 i}{3\sqrt{2}}\delta_{\ell 1}- 
  \frac{i^{\ell}}{\sqrt{2}\ell(\ell+1)}$ & 0
 \\  \hline
  $VB$ & 0& $\frac{1}{2}\left(-\frac{i}{3} \delta_{\ell1} + \frac{i^\ell}{\sqrt{2\ell (\ell+1)}}\right)$ & 0 & $\frac{i}{3\sqrt{2}}\delta_{\ell 1}- 
  \frac{i^{\ell}}{\sqrt{2}\ell(\ell+1)}$
        \\  \hline
  $TE$  &
        $\frac{i^{\ell}}{2} N_\ell^{-1}$ &0& $-i^\ell \frac{N_\ell^{-1}}{\sqrt{\ell(\ell+1)}}$ & 0 \\  \hline
  $TB$ & 0& $\frac{i^{\ell}}{2} N_\ell^{-1}$
          & 0 & $-i^\ell \frac{N_\ell^{-1}}{\sqrt{\ell(\ell+1)}}$ \\ \hline     
\end{tabular}
\caption{
The factors $F^{{\cal O},X}_\ell$ for spin-1 and spin-2 TAM waves in the distant-source limit ($kr_s \to \infty$).}
\label{tab:F}
\end{table*}

\subsection{Timing-residual--birefringence cross correlation}

To characterize chiral GWs, we express the right/left-handed GWs as
\begin{align}
    h^{k,\alpha,\pm}_{(\ell m)} = \frac{1}{\sqrt{2}}\left(h^{k,\alpha E}_{(\ell m)} \mp i h^{k,\alpha B}_{(\ell m)} \right).
\end{align}
We introduce the chirality parameters $\Delta \chi_\alpha$ through $P_{\alpha,\pm}(k) = (1 \mp \Delta \chi_\alpha)P_{\alpha}(k)$ with $\alpha \in \{T,V\}$.
Note that the power spectrum is related to the TAM modes as 
\begin{align}
\vev{ h_{(\ell m)}^{k,\alpha E} \left(h_{(\ell'm')}^{k',\alpha E} \right)^* } &= \vev{ h_{(\ell m)}^{k,\alpha B} \left(h_{(\ell'm')}^{k',\alpha B} \right)^* } \nonumber \\
&= \frac{(2\pi)^3}{k^2} \delta_{\ell \ell'} \delta_{mm'} \delta(k-k') P_\alpha(k), \\
\vev{ h_{(\ell m)}^{k,\alpha, \pm} \left(h_{(\ell'm')}^{k',\alpha, \pm} \right)^* } &= \frac{(2\pi)^3}{k^2} \delta_{\ell \ell'} \delta_{mm'} \delta(k-k') P_{\alpha, \pm}(k), \\
\vev{ h_{(\ell m)}^{k,\alpha E} \left(h_{(\ell' m')}^{k',\alpha B} \right)^* } \nonumber \\
= i \frac{(2\pi)^3}{k^2} &\delta_{\ell \ell'} \delta_{mm'} \delta(k-k') P^{(\alpha E,\alpha B)}(k).
\label{eq:h_tam_eb}
\end{align}
Then, we find 
\begin{equation}
     {P}^{(\alpha E,\alpha B)}(k) = \Delta \chi_\alpha P_\alpha(k).
\end{equation}
$\Delta \chi_\alpha$ describes the degree of parity breaking. 
For example, $\Delta \chi_\alpha = \pm 1$ corresponds to the maximum parity breaking.

Although we have obtained the expressions for one TAM mode of tensor perturbation (Eqs.~(\ref{eq:z_final}) and (\ref{eq:psi_final})), we can use them as those for one frequency mode by replacing $k \to f$.
This replacement $k \to f$ is justified within the TAM formalism for massless gravitons.
In fact, the factor $\ee^{-ikt}$ in Eqs.~(\ref{eq:z_2}) and (\ref{eq:psi_tam}) comes from the relation $f = k$ for massless gravitons.
Then, we can easily see (suppressing the $k$, or frequency, dependence) that the angular two-point correlation functions satisfy
\begin{align}
  \vev{z(\bm{\hat n}) \psi^*(\bm{\hat m})} = \begin{cases}
  i\Delta \chi_T \vev{z(\bm{\hat n}) z^*(\bm{\hat m})}  &(\text{spin-}2) \\
  i\cfrac{\Delta \chi_V}{2} \vev{z(\bm{\hat n}) z^*(\bm{\hat m})}  &(\text{spin-}1)
  \end{cases}.
\end{align}
We thus verify the observation of Ref.~\cite{Liang:2025vji} that the $z\psi$ cross-correlation has the same (Hellings-Downs) angular dependence as the $zz$ auto-correlation.  For the spin-1 cases, the two correlation functions also have the same angular dependence, which in this case is given by $\vev{z(\bm{\hat n}) \psi^*(\bm{\hat m})} \propto -2 \ln [\sin(\Theta/2)] - 1- (4/3) \cos\Theta$ with $\cos\Theta = \bm {\hat n} \cdot \bm{\hat m}$~\cite{LeeJenetPrice:2008,Mihaylov:2018uqm,Qin:2018yhy}. 

\subsection{Astrometry}

\subsubsection{Harmonic analysis}
A GW background will also produce subtle oscillations in the angular locations of sources \cite{Braginsky:1989pv,Book:2010pf}.  The pattern of angular deflections $(\delta \hat n)_a(\bm{\hat n})$ as a function of position on the sky can be expanded (see, e.g., Ref.~\cite{Qin:2018yhy})
\begin{equation}
    (\delta \hat n)_a(\bm {\hat n}) = \sum_{\ell m} \left[ E_{\ell m} Y_{(\ell m)a}^{E}(\bm{\hat n}) + B_{\ell m} Y_{(\ell m)a}^{B}(\bm{\hat n}) \right],
\end{equation}
in terms of even- and odd-parity vector spherical harmonics $Y_{(\ell m)a}^E(\bm{\hat n})$ and $Y_{(\ell m)a}^B(\bm{\hat n})$, respectively.  In the absence of parity breaking, there will be a cross-correlation between $\psi$ and $B$. On the other hand, a chiral GW background will induce a cross-correlation between $\psi$ and $E$.  Given the parallels noted above between the angular-deflection field $\psi(\bm{\hat n})$ and timing-residual field $z(\bm{\hat n})$, and following Ref.~\cite{Qin:2018yhy}, the parity-conserving $\psi$-deflection cross-correlation will, in harmonic space, be
\begin{eqnarray}
     C_\ell^{\psi B,X} &=& 32 \pi^2 F_\ell^{\psi,X} \left(F_\ell^{B,X}\right)^* \nonumber \\
     & & \times \int\, \dd f\, \frac{6 H_0^2 \Omega_X(f)}{2 (2\pi)^3 f^3 } W_E(f) W_\psi^*(f),
\end{eqnarray}
for $X=TB$ or $VB$ with $F_\ell^{B,X}$ as given in the Table~\ref{tab:F} (from Ref.~\cite{Qin:2018yhy}).  
$H_0$ is the Hubble constant, $\Omega_X$ is the energy density parameter for $X$-mode GWs, and $W_\beta$ ($\beta \in \{E,\psi\}$) is the window function associated with the cadence of observations (note $W_B = W_E$).
From $F^{E,TE} = F^{B,TB}$, we can express the cross-correlation between $\psi$ and the E-mode due to spin-2 GWs as 
\begin{align}
    \label{eq:c_psi_e_t}
C_\ell^{\psi E,T} &= i\Delta \chi_T C_\ell^{\psi B,TB} = i\Delta \chi_T C_\ell^{z E,TE}.
\end{align}
On the other hand, for spin-1 GWs, we find 
\begin{align}
    \label{eq:c_psi_e_v}
    C_\ell^{\psi E,V} = i\Delta \chi_V (1-2\delta_{\ell 1} ) C_\ell^{\psi B,VB} = i\frac{\Delta \chi_V}{2} C_\ell^{z E,VE}, 
\end{align}
where the factor $(1-2\delta_{\ell 1})$ comes from the difference between $F^{E,VE}_\ell$ and $F^{B,VB}_\ell$.

\subsubsection{Configuration space}

Parity-even and parity-odd astrometry angular two-point correlations are defined by first defining components $(\delta n)^\parallel$ and $(\delta n)^\perp$ of the deflection that are, respectively, parallel and perpendicular to the great arc connecting the two points being correlated.  
Generalizing the results in Ref.~\cite{Qin:2018yhy}, we obtain the parity-conserving $\psi$-deflection cross-correlation in configuration space as 
\begin{equation}
     \vev{\psi(\bm{\hat n}) (\delta n)^\perp(\bm{\hat m})}  = \sum_\ell \frac{2\ell+1}{4\pi} \frac{C_\ell^{\psi B} P^1_{\ell}(\bm{\hat n}\cdot \bm{\hat m})}{ \sqrt{\ell(\ell+1)}},
\end{equation}
in terms of associated Legendre polynomials $P^m_{\ell}(x)$.  The sum converges fairly quickly and is easily evaluated, and the result is shown (for both spin-1 and spin-2) in Fig.~4 of Ref.~\cite{Qin:2018yhy}.  
Similarly, we obtain the parity violation contribution as 
\begin{equation}
     \vev{\psi(\bm{\hat n}) (\delta n)^\|(\bm{\hat m})}  = \sum_\ell \frac{2\ell+1}{4\pi} \frac{C_\ell^{\psi E} P^1_{\ell}(\bm{\hat n}\cdot \bm{\hat m})}{ \sqrt{\ell(\ell+1)}}.
\end{equation}
Also, from Eqs.~(\ref{eq:c_psi_e_t}) and (\ref{eq:c_psi_e_v}), we can easily see 
\begin{align}
  \vev{\psi(\bm{\hat n}) (\delta n)^\|(\bm{\hat m})} = \begin{cases}
  i\Delta \chi_T \vev{z(\bm{\hat n})  (\delta n)^\|(\bm{\hat m}}  &(\text{spin-}2) \\
  i\cfrac{\Delta \chi_V}{2} \vev{z(\bm{\hat n})  (\delta n)^\|(\bm{\hat m})}  &(\text{spin-}1)
  \end{cases}.
\end{align}

\section{Conclusions}

We have built upon the suggestion of Ref.~\cite{Liang:2025vji} to use the rotation of the linear polarization of electromagnetic waves from pulsars, in cross-correlation with pulsar timing.  The cross-correlation of this rotation, a pseudoscalar on the sky, with the timing residual, a scalar, allows a probe of the chirality of the GW background.

Here we have re-derived the cross-correlation using the TAM formalism and then also generalized (using results from Ref.~\cite{Qin:2018yhy}) to cross-correlations, induced by a parity-conserving or chiral GW background, with astrometry observables as well.  We have provided results for both the standard spin-2 GWs that arise in general relativity as well as the spin-1 GWs that may arise in alternative-gravity theories. We have here assumed the GW background to be isotropic, but we note that the generalization to correlations involving electromagnetic wave birefringence in the presence of anisotropies (and/or GW linear polarization) can easily be done using the techniques of Refs.~\cite{Bernardo:2023jhs,AnilKumar:2023yfw, AnilKumar:2023hza,Inomata:2024kzr}.

\begin{acknowledgments}
We thank Qiuyue Liang, Kimihiro Nomura, and Hidetoshi Omiya for useful comments on our manuscript.
This work was supported by NSF Grant No.\ 2412361,
NASA ATP Grant No.\ 80NSSC24K1226, and the Templeton Foundation.
\end{acknowledgments}
 
%%%%%%%%%%%%%%%%%%%%%%%%%%%%%%%%%%%
\appendix
\section{Relations between the derivative of TAM modes and spherical harmonics}
\label{app:some_rel}

In this Appendix, we summarize useful relations between the derivative of TAM modes and spherical harmonics.
From Eq.~(38) in Ref.~\cite{Dai:2012bc}, we have
\begin{align}
  Y^B_{(\ell m)a}(\bm{\hat n}) = \frac{-r}{\sqrt{\ell (\ell+1)}} \varepsilon_{abc} \hat n^b \partial^c Y_{(\ell m)}(\bm{\hat n}).
\end{align}
From Eq.~(A9) in Ref.~\cite{Qin:2018yhy}, we have
\begin{align}
  \hat n^a \Psi^{Y}_{(\ell m)ab}(\bm x) = R^{B,Y}_\ell(kr) Y^B_{(\ell m)b}(\bm {\hat n}),
\end{align}
where $Y \in\{TB,VB\}$. 
Combining these, we obtain 
\begin{align}
  &\varepsilon^{ab}_{\ \ c} \hat n^c \hat n^d \partial_a \Psi^{Y}_{(\ell m)bd}(\bm x) \nonumber  \\ 
 &= \varepsilon^{ab}_{\ \ c} \hat n^c \partial_a  R^{B,Y}_\ell(kr) Y^B_{(\ell m)b}(\bm {\hat n}) \nonumber \\  
  &= \frac{-r}{\sqrt{\ell (\ell+1)}} R^{B,Y}_\ell(kr) \nabla^2 Y_{(\ell m)}(\bm{\hat n}) \nonumber \\
  &= \frac{\sqrt{\ell(\ell+1)}}{r} R^{B,Y}_\ell(kr) Y_{(\ell m)}(\bm{\hat n}),
  \label{eq:tb_rel}
\end{align}
where we have used $\varepsilon^{a}_{\ bc} \varepsilon_{ade} = \delta_{bd}\delta_{ce} - \delta_{be}\delta_{cd}$ and $r^2 \nabla^2 Y_{(\ell m)} = -\ell (\ell+1) Y_{(\ell m)}$.

%%%%%%%%%%%%%%%%%%%%%%%%%%%%%%%%%%%
%%%%%%%%%%%%%%%%%%%%%%%%%%%%%%%%%%%
%%%%%%%%%%%%%%%%%%%%%%%%%%%%%%%%%%%
\small
\bibliography{pulsar_polarimetry}{}

\end{document}